\documentclass[runningheads]{llncs}
\usepackage{graphicx}
\usepackage{amsmath}
\usepackage{url}
\usepackage[table,x11names]{xcolor}
\definecolor{mygray}{gray}{0.9}
\definecolor{mygray_light}{gray}{0.95}
\usepackage{floatrow}
\newfloatcommand{capbtabbox}{table}[][\FBwidth]

\begin{document}
\title{Learning from Thresholds: Fully Automated Classification of Tumor Infiltrating Lymphocytes for Multiple Cancer Types}
\titlerunning{ }
%
\author{Shahira
Abousamra , Le Hou, Rajarsi Gupta, Chao Chen, Dimitris Samaras, Tahsin Kurc , Rebecca Batiste, Tianhao Zhao, Shroyer Kenneth, Joel Saltz}
\authorrunning{ }
%
\institute{Stony Brook University, NY, USA
}
\maketitle              
\begin{abstract}

Deep learning classifiers for characterization of whole slide tissue morphology require large volumes of annotated data to learn variations across different tissue and cancer types. As is well known, manual generation of digital pathology training data is time consuming and expensive. In this paper, we propose a semi-automated method for annotating a group of similar instances at once,  instead of collecting only per-instance manual annotations. This allows for a much larger training set, that reflects visual variability across multiple cancer types and thus training of a single network which can be automatically applied to each cancer type without human adjustment. We apply our method to the important task of classifying Tumor Infiltrating Lymphocytes (TILs) in H\&E images. Prior approaches were trained for individual cancer types,  with smaller training sets and human-in-the-loop threshold adjustment. We utilize these thresholded results as large scale ``semi-automatic'' annotations. Combined with existing manual annotations, our trained deep networks are able to automatically produce better TIL prediction results in 12 cancer types, compared to the  human-in-the-loop approach.


\end{abstract}
\vspace{-0.5cm}
\section{Introduction}
\label{sec:intro}
\begin{figure}
\centering
\vspace{-0.5cm}
\includegraphics[width=0.85\textwidth]{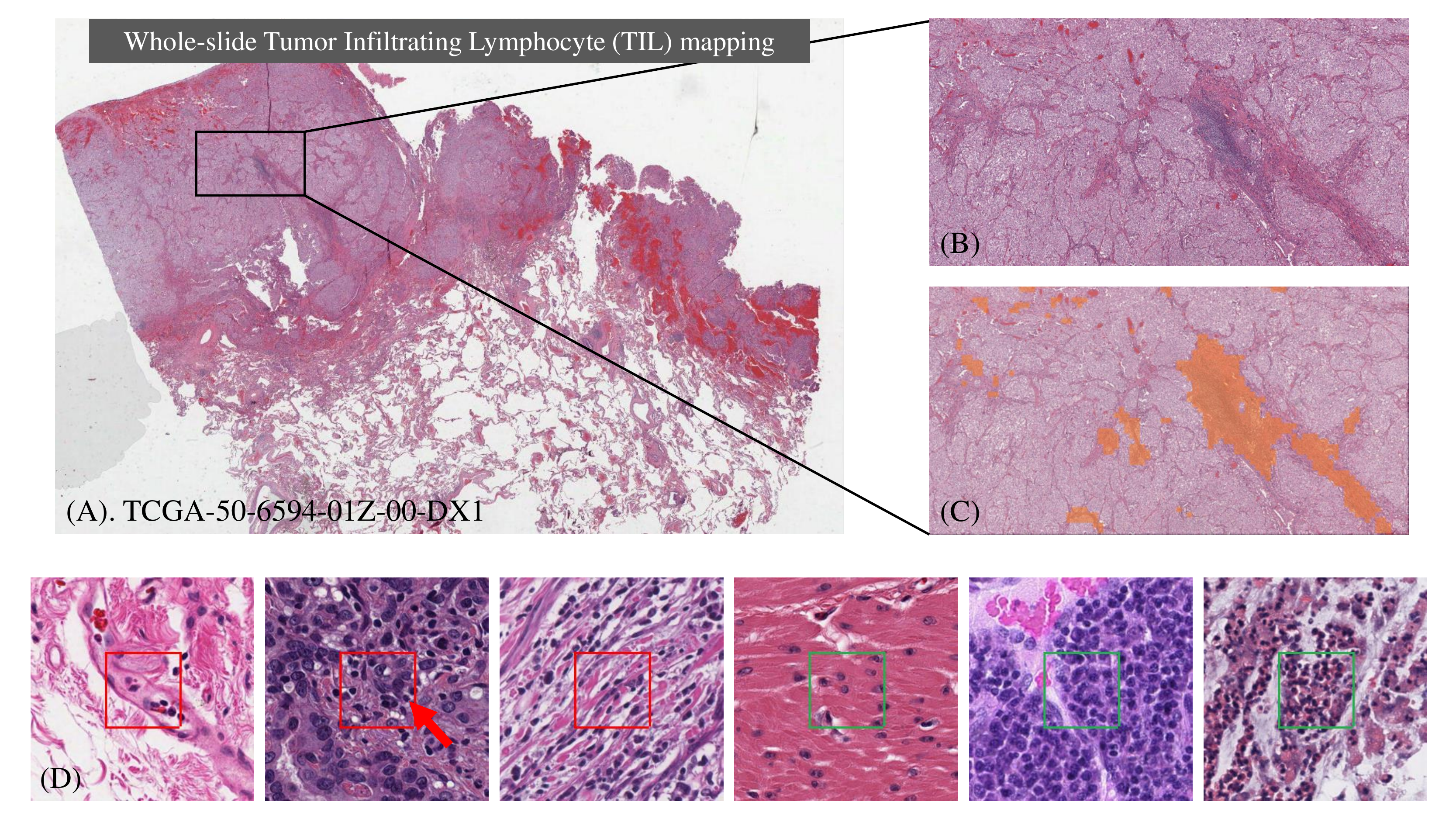}
\vspace{-0.2cm}
\caption{The problem of identifying Tumor Infiltrating Lymphocyte (TIL) regions in gigapixel pathology WSIs of 12 cancer types. \textbf{(A).} H\&E stained WSI of lung adenocarcinoma. \textbf{(B).} Example of a region of tissue. \textbf{(C).} Example of a thresholded TIL map overlaid on the region of tissue. \textbf{(D).} Examples of TIL positive (framed in red) and negative (framed in green) patches. A lymphocyte is typically dark, round to ovoid, and relatively small compared to tumor and normal cells (right arrow). Also note the wide spectrum of heterogeneity in pathology images.}
\label{fig:first_fig}
\end{figure}

\vspace{-0.1cm}
Quantitative characterization of tumor infiltrating lymphocytes (TILs) is becoming rapidly important in precision medicine \cite{SALTZ2018181,barnes2018development,corredor2018watershed,steele2018measuring,thorsson2018immune,mahmoud2011tumor,salgado2014evaluation,hendry2017assessing,john2016assessing}. With the growth of cancer immunotherapy, these characterizations can provide clinically significant information to further our understanding of the immune response in cancer patients across a wide spectrum of cancer types. For instance, high densities of TILs are shown to correlate with favorable clinical outcomes \cite{mlecnik2011tumor} that include longer disease-free survival (DFS) and/or improved overall survival (OS) in multiple cancer types \cite{angell2013immune}. Recent studies also suggest that the spatial context of immune response within the tumor microenvironment is important in cancer prognosis~\cite{mlecnik2011histopathologic}.
In this work, we target the problem of identifying Tumor Infiltrating Lymphocytes (TILs) in Whole Slide Images (WSIs) obtained from multiple cancer types. In particular, we partition a WSI into patches and classify each image patch as TIL positive or TIL negative, as shown in Fig. \ref{fig:first_fig}.

An earlier work~\cite{SALTZ2018181} developed a deep 
learning pipeline that classifies image patches of 
50$\times$50 square microns into TIL positive and 
TIL negative classes. 
%
%
In order to train neural network models that generalize across multiple cancer types,
the authors collected manually labeled individual patches in an 
iterative train-review-retrain process, which is labor intensive and time consuming.
Nevertheless, the model trained using only strong annotations cannot generalize across cancer types without 
a human-in-the-loop, TIL probability thresholding step, 
which compensates for potential image-specific model bias (model may over- 
or under-estimate TILs). 
In this step the human expert chooses and applies thresholds over the model's probability output to correct 
for the prediction bias in given set of WSIs to an extent. Using this approach, they generated {\em TIL maps} 
for roughly 5,000 H\&E WSIs from the Cancer Genome Atlas (TCGA) repository, which were utilized in a variety of downstream analyses 
to correlate the spatial features of the TIL maps with molecular features in a Pan Cancer 
study~\cite{SALTZ2018181,thorsson2018immune}.  

\vspace{-0.5cm}


Most research on TIL prediction focuses on a specific cancer type, such as in \cite{8104187} and \cite{Linder157}. 
The existing state-of-the-art approach \cite{SALTZ2018181} spent months collecting per-instance manual annotations, in order to train the model to generalize across cancer types. In this paper, we propose to use human thresholds applied on groups of instances as weak annotations, in addition to per-instance strong (manual) annotations. These weak, semi-automatic annotations are extremely cost effective - thresholding a TIL map costs only several minutes, but generates typically 100K annotations. We demonstrate that \textbf{we are able to utilize TIL map thresholds as weak annotations, and train well-engineered neural networks to achieve high quality TIL prediction results across multiple cancer types without needing any human thresholding.} This goal could not be achieved by the existing approach \cite{SALTZ2018181} despite months of manually annotating images.
Our contributions are as follows:
\begin{enumerate}
    \item We advocate for the methodology of semi-automatic annotation via thresholding a possibly biased classifier. Using this method, one is able to collect large scale annotated datasets across cancer types.
    \item We will release  software (upon publication) for fully automated classification of TILs in WSIs of 12 cancer types that has the following advantages.
    \begin{enumerate}
        \item Our TIL classification model generates TIL maps with similar or better quality compared to the state-of-the-art TIL pipeline described in \cite{SALTZ2018181}.
        \item Our models do not require any human adjustment, whereas the current \cite{SALTZ2018181} TIL pipeline requires manual threshold adjustment using a specific software stack. Our approach makes it possible to obtain equivalent or better results in a much more scalable manner. 
    \end{enumerate} 
    \item In addition  to  making our models and Tensorflow CNN codes available,  we will also release approximately 10,000 new TIL maps. We will release two new TIL maps per image - one from our VGG-16 model and one from our Inception v4 model (see Section \ref{sec:deep-neural-network-models}) - corresponding to each of the approximately 5,000 WSI/TIL map pairs that are currently available in the Cancer Imaging Archive (TCIA).
\end{enumerate}


\vspace{-0.4cm}
\section{Our methods}
\vspace{-0.1cm}
Lymphocytes are a specific type of immune cell that are classified as B- or T-cells, where plasma cells are the terminally differentiated form of B-lymphocytes. However, B- and T-cells are indistinguishable in H\&E sections. Therefore, a given 50$\times$50 square micron tissue patch is considered Tumor Infiltrating Lymphocyte (TIL) positive if it contains a significant number of lymphocytes and/or plasma cells \cite{SALTZ2018181}. Examples of positive and negative patches are shown in Fig. \ref{fig:first_fig}. 
TIL maps for WSIs are the patch-by-patch (non-overlapping) TIL classification results for each WSI.

Our goal is to train TIL classification models that generalize across 12 cancer types: Urothelial carcinoma of the bladder (BLCA), Invasive carcinoma of the breast (BRCA), Cervical squamous cell carcinoma and endocervical adenocarcinoma (CESC), Colon adenocarcinoma (COAD), Lung adenocarcinoma (LUAD), Lung squamous cell carcinoma (LUSC), Pancreatic adenocarcinoma (PAAD), Prostate adenocarcinoma (PRAD), Rectal adenocarcinoma (READ), Skin Cutaneous Melanoma (SKCM), Stomach adenocarcinoma (STAD), Endometrial Carcinoma of the Uterine Corpua (UCEC). All of the WSIs utilized in the study come from the TCGA \cite{tcga}. We do not predict TILs in Uveal Melanoma (UVM) in the existing approach \cite{SALTZ2018181} due to the paucity of TILs in this type of cancer.

\vspace{-0.3cm}
\subsection{Manual and semi-automatic annotations}
\label{sec:training-data-collection}
All training patches used in our work are in 100$\times$100 pixel resolution, 20X magnification, and are annotated as TIL positive or TIL negative.

\vspace{-0.35cm}
\subsubsection{Manually annotated 86K patches}
As described in Section \ref{sec:intro}, the human-in-the-loop TIL classification baseline \cite{SALTZ2018181} collected image patches with human annotations in a cumulative fashion, for seven cancer types (BRCA, COAD, LUAD, PAAD, PRAD, SKCM, UCEC with 2.9K, 4.0K, 32K, 1.9K, 5.5K, 34K, 5.4K patches respectively). As a result, the existing approach used a subset of those training patches for the generation of TIL maps for each cancer type. In this work, we utilize the entire set of 86K patches for training. There are 64,381 TIL negative patches and 21,773 TIL positive patches.

\vspace{-0.35cm}
\subsubsection{Semi-automatically annotated 301K patches}
The human-in-the-loop TIL classification baseline \cite{SALTZ2018181} generated over 5,000 TIL maps by applying the TIL classification network on WSIs, followed by a manual thresholding step over predicted TIL probability maps. These thresholded TIL maps have not been used as training data before our work. We extracted an average of 120 patches per WSI for 2,500 WSIs, based on patient IDs. The rest of the patients IDs were reserved for testing. This approach generated 301K annotated patches. We consider these patches as ``semi-automatically annotated".

\vspace{-0.35cm}
\subsubsection{Mixture of manually and semi-automatically annotated patches}
\textit{For cancer types that have manual annotations, we use manual annotations. For cancer types that do not have manual annotations, we use semi-automatic annotations}, except that we did not use any annotation from BLCA WSIs, since TILs in BLCA are relatively easy to identify. The resulting training set contains 86K manually annotated patches and 69K semi-automatically annotated patches. 

\vspace{-0.35cm}
\subsection{Deep neural network models}
\label{sec:deep-neural-network-models}
\vspace{-0.1cm}
For the task of TIL mapping in gigapixel WSIs, we selected neural networks by considering both their computational complexity and capacity. As a result, we used two CNN architectures: the VGG 16-layer network \cite{Simonyan14c-vgg}, and Inception-V4  \cite{szegedy2017inception}. The VGG-16 and Inception-V4 networks are pretrained on ImageNet. 
In order to match input dimensions, we resized the input 100 $\times$ 100 pixel patches to 224 $\times$ 224 for VGG-16 and 299 $\times$ 299 pixels for Inception-V4, respectively. Given an input image patch, a network outputs the probability of the input being TIL positive.

We used a sigmoid as the activation function in the last layer. We trained both networks with the Adam optimizer using an initial learning rate of 0.0005. We used a batch size of 128 during training. We did not use batch normalization, since  it did not improve performance in this task, mainly due to the imbalance in the training dataset. We applied three types of stochastic data augmentation: (1). Shifting input patches left/right and up/down by a random number of pixels in the range of $[-20, +20]$; (2). Rotating and flipping input patches; (3). Color augmentation by small variations to brightness and color in the HSL space.

\vspace{-0.35cm}
\section{Experiments}
\vspace{-0.1cm}

\vspace{-0.1cm}
\subsubsection{Thresholding probability output}
The final binary prediction (TIL positive or negative) is obtained by thresholding the probability output of the network. For our methods, the threshold for each network is a fixed value, selected using a validation dataset with 652 LUAD patches. In particular, we select the threshold to minimize the difference between false positive and false negative rates based on the ROC curve, which we refer to as Youden's index threshold \cite{youden:1950}.

\vspace{-0.2cm}
\begin{table}[H]
\vspace{-0.2cm}
\caption{The list of models we compare in our experiments}
\begin{floatrow}
\resizebox{1\textwidth}{!}{%
\begin{tabular}{ |c|c| } 
 \hline
 \rowcolor{mygray} 
 	Model Name & Training set (see Section \ref{sec:training-data-collection})  \\ 
 \hline
 Baseline-Youd \& Baseline-HITL \cite{SALTZ2018181,hou2019sparse}  & Subsets (at most 52K) of 86K manually annotated \\
 \hline
 VGG-manual \& Incep-manual & 86K manually annotated \\
 \hline
 VGG-semi \& Incep-semi & 301K semi-automatically annotated \\
 \hline
 VGG-all \& Incep-all & 301K semi-automatically + 86K manually annotated \\
 \hline
 VGG-mix \& Incep-mix & 69K semi-automatically + 86K manually annotated \\
 \hline
\end{tabular}
}
\end{floatrow}
\label{table:models-config}
\vspace{-0.2cm}
\end{table}

\vspace{-1.1cm}
\begin{table}
\resizebox{1\textwidth}{!}{
\vspace{-0.4cm}
 \begin{tabular}{ |c|c|c|c|c|c|c|c|c|c| } 
 \hline
 \rowcolor{mygray}
  	 & \multicolumn{3}{c|}{Overall} & \multicolumn{3}{c|}{LUAD} & \multicolumn{3}{c|}{BRCA} \\ \cline{2-10}
 \rowcolor{mygray_light} 
 	Model Name & F1-score & Accuracy & AUC & F1-score & Accuracy & AUC & F1-score & Accuracy & AUC \\
 \hline
 Baseline-HITL & 0.85 & 79.56\% & 0.798 & 0.78 & 73.6\% & 0.772 & 0.77 & 74.9\% & 0.808 \\
 \hline
 Baseline-Youd & 0.85 & 78.22\% & 0.798 & 0.79 & 72.9\%  & 0.772  & 0.76 & 71.9\% & 0.808 \\
 \hline
 VGG-manual  & 0.87 & 82.44\% & 0.899 & 0.87 & 83.3\% & 0.886 & 0.88 & 87.5\% &  0.948 \\
 \hline
 Incep-manual & 0.87 & 82.44\% & 0.890 &  0.86 & 83.3\%  & 0.894 & 0.88 & 86.9\% &  0.936 \\
 \hline
 VGG-semi  & 0.85 & 80.33\% &  0.870 &  0.83 & 79.26\% & 0.872 & 0.86 & 84.40\% & 0.930 \\
 \hline 
 Incep-semi  & 0.84  & 78.56\% &  0.866 & 0.84 & 80.27\% & 0.871 & 0.86 &  84.40\% & 0.926 \\
 \hline 
 VGG-all  & 0.87 & 81.67\% & 0.878  & 0.85 & 81.27\% & 0.879 &  0.86 & 84.10\% & 0.933 \\
 \hline 
 Incep-all  & 0.87  & 81.89\% & 0.885  & 0.86 & 81.94\% & 0.884 & 0.86 & 83.49\% & 0.941 \\
 \hline  
 VGG-mix  & \textbf{0.88} & \textbf{84.22\%} & 0.9  & 0.87 & 84.28\% & 0.902 &  \textbf{0.91} & \textbf{90.21\%} & \textbf{0.953} \\
 \hline 
 Incep-mix  & \textbf{0.88}  & 83.22\% & \textbf{0.903}  & \textbf{0.88} & \textbf{84.95\%} & \textbf{0.914} & 0.9 & 88.69\% & \textbf{0.953} \\
 \hline  
\end{tabular}
}
\vspace{-0.2cm}
\caption{\label{tab:patch-evaluation} Results on 12 cancer types (overall), and for LUAD and BRCA specifically. Networks trained with a mixture of manual and semi-automatic annotations (VGG-mix, Incep-mix) outperform other methods consistently.}%
\end{table}
\vspace{-0.4cm}

\vspace{-0.7cm}
\subsubsection{List of tested models}
As described in Section \ref{sec:training-data-collection} and \ref{sec:deep-neural-network-models}, we have three training datasets and two network architectures. We tested combinations of networks and datasets and named the models in Table \ref{table:models-config}. For the human-in-the-loop baseline method \cite{SALTZ2018181}, we directly used two released versions of TIL predictions:
\begin{enumerate}
    \item Gray-scale TIL maps as probability predictions. We applied a constant threshold, 0.26, that was selected by using Youden's method as described above. We call this method \textbf{Baseline-Youd}. Note that the original approach does not use a fixed threshold: it only uses human-adjusted thresholds.
    \item Binary prediction results given by the human-in-the-loop thresholding step. We call this method \textbf{Baseline-HITL}.
\end{enumerate}
\vspace{-0.1cm}
We refer to the VGG 16-layer network as \textbf{VGG} and the Inception-V4 network as \textbf{Incep}. Youden's index threshold for our VGG/Incep-mix is 0.42/0.10.

\begin{figure}
\centering
\includegraphics[trim={0 1cm 0 0},clip,width=1.0\textwidth]{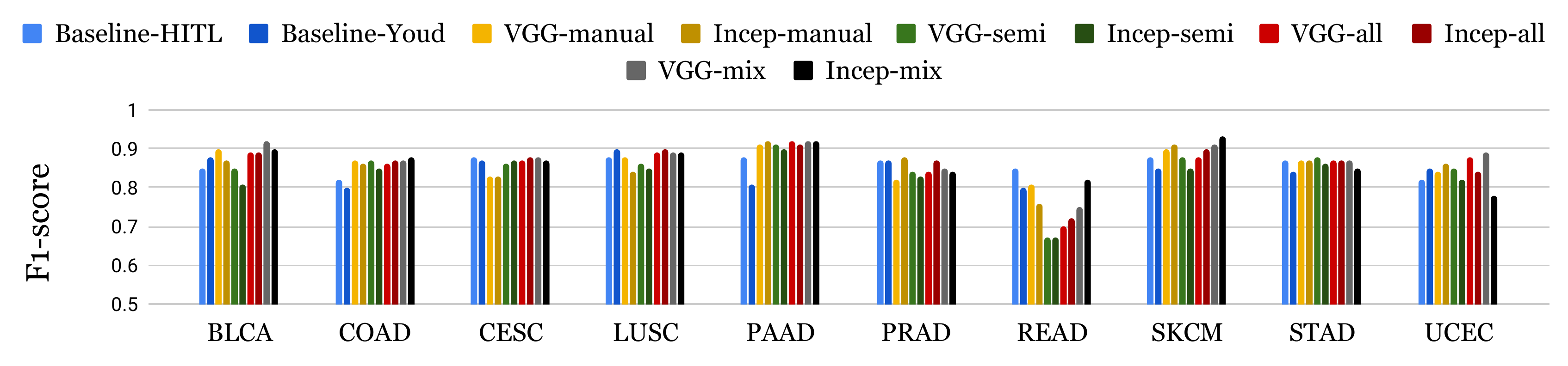}
\includegraphics[trim={0 1cm 0 0},clip,width=1.0\textwidth]{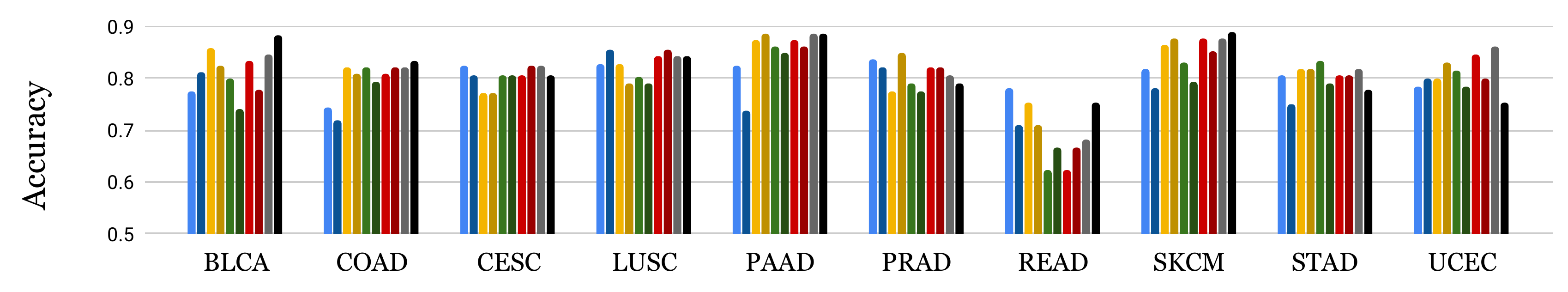}
\includegraphics[width=1.0\textwidth]{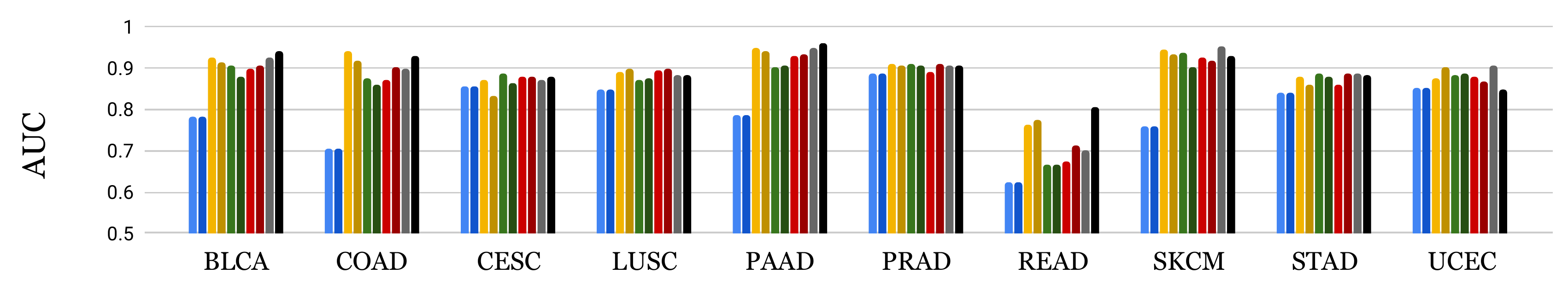}
\vspace{-0.6cm}
\caption{{\label{fig:patch-evaluation-other-cancer}}Detailed patch classification results on the remaining 10 cancer types (results on LUAD and BRCA are in Table \ref{tab:patch-evaluation}). There are 65 to 85 test patches in each cancer type. The proposed VGG-mix and VGG-mix perform well across 10 cancer types. Adding semi-automatic annotations in cancer types with manual annotations (COAD, PAAD, PRAD, SKCM, UCEC) does not improve the average scores in these cancer types. Adding semi-automatic annotations only in cancer types without manual annotations (CESC, LUSC, READ, STAD) indeed improves their average scores.}
\end{figure}

\vspace{-0.3cm}
\subsection{Accuracy of patch classification}
\vspace{-0.1cm}
We evaluated the performance of the methods listed in Table \ref{table:models-config} in classifying individual patches. We manually annotated a test set of 900 patches from all  12 cancer types. For each cancer type,  roughly  the same number of patches is available. In order to evaluate the performance of the TIL classification models for individual cancer types, we annotated \textit{additional} 457 patches for two of the most common types of cancer: LUAD and BRCA. The test sets were created according to the TIL probability maps released by the baseline work \cite{SALTZ2018181}, where test patches were sampled across the range of predicted TIL probabilities. The test set that we created is expected to be more challenging in comparison to the test set that was used in the baseline model \cite{SALTZ2018181}, where test patches were sampled differently. We use the F1-score, accuracy, and Area Under the ROC curve (AUC) as evaluation metrics. The overall results in 12 cancer types, the LUAD test set, and the BRCA test set are shown in Table \ref{tab:patch-evaluation}. We include results for the remaining 10 cancer types  in Fig. \ref{fig:patch-evaluation-other-cancer}.

From the results, we conclude that networks trained with the proposed annotations (VGG-mix, Incep-mix) achieve the best performance. Moreover, our fully automated method outperforms the baseline method with human adjustment (Baseline-HITL). Note that the original baseline method \cite{SALTZ2018181} did not use Youden's index threshold - it only used manually adjusted thresholds. Using a default threshold of 0.5 instead of Youden's index threshold, the accuracy of the baseline drops from 78.22\% to 70.22\% and the F1-score drops from 0.85 to 0.76.

\vspace{-0.3cm}
\begin{figure}
\centering
\includegraphics[width=1.0\textwidth]{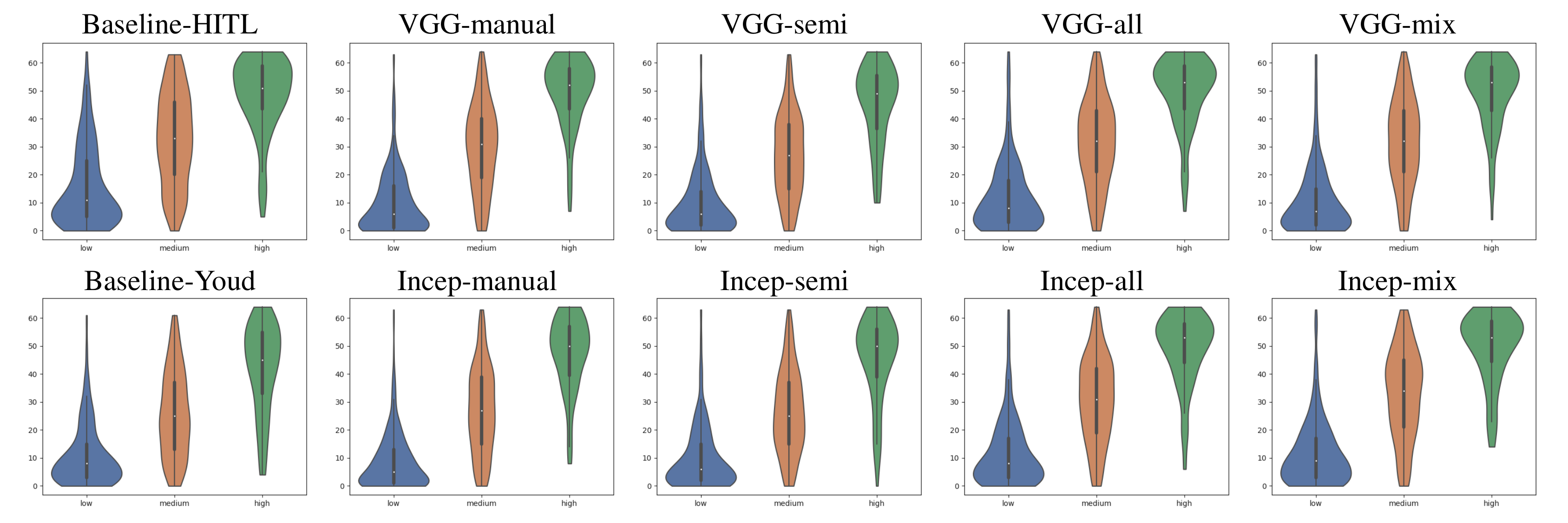}
\caption{\label{fig:violin} Performance of various models on identifying regions with low to high TILs. x-axis: ground truth labels of low/medium/high TILs; y-axis: TIL prediction results.}
\end{figure}

\vspace{-0.1cm}
\subsection{Identifying regions with low/medium/high TILs}
\vspace{-0.1cm}
 We also evaluated the performance of these models in terms of identifying the amount of lymphocytes in tissue regions using a binary TIL classifier. For this purpose, three pathologists labeled large patches of $800\times800$ pixels as low, medium, or high with respect to the amount of TILs within the area of tissue in the large patch. We then took the rounded average of multiple pathologists' ratings as the final label. In order to obtain the automated prediction results, the large patches are divided into 64 patches (each $100\times100$ pixels) for input into the model. We counted the number of predicted TIL positive patches as the prediction results of the model, which ranges from 0 to 64. We plot the scores for each category in violin plots, as shown in Fig. \ref{fig:violin}. Networks trained with proposed annotation methods, VGG-mix and Incep-mix, distinguish regions with low/medium/high TILs better than the baseline method \cite{SALTZ2018181}.

\vspace{-0.3cm}
\section{Conclusions}
\vspace{-0.1cm}
The task of identifying TILs in pathology WSIs is becoming increasingly important as more immunotherapy options continue to emerge. In this paper, we propose to use manually thresholded TIL maps given by a possibly biased neural network as semi-automatic annotations to augment our training data to extensively train larger neural networks. Neural networks trained with both manual and semi-automatic annotations perform the best, especially on cancer types that do not have manual annotations. In comparison to the  state-of-the-art, human-in-the-loop method \cite{SALTZ2018181}, our model achieves better TIL classification results in a much more scalable manner that can be implemented in future studies with large collections of WSIs. We will release these models, code, and roughly 10,000 TIL maps of 5,000 WSIs from our VGG-16 and Inception-V4 networks.

\vspace{-0.3cm}
\bibliographystyle{splncs04}
\bibliography{references}

\end{document}